\documentclass[apj]{emulateapj}

\begin{document}

\shorttitle{X-rays from XSS J12270--4859}
\shortauthors{Bogdanov et al.}

\title{X-ray Observations of XSS J12270--4859 in a New Low State:\\ A Transformation to a Disk-Free Rotation-Powered Pulsar Binary}

\author{Slavko Bogdanov\altaffilmark{1}, Alessandro
  Patruno\altaffilmark{2,3}, Anne M.~Archibald\altaffilmark{3}, Cees
  Bassa\altaffilmark{3},\\ Jason W.~T.~Hessels\altaffilmark{3,4},
  Gemma H.~Janssen\altaffilmark{3}, Ben W.~Stappers\altaffilmark{5} }

\altaffiltext{1}{Columbia Astrophysics Laboratory, Columbia University, 550 West 120th Street, New York, NY 10027, USA; slavko@astro.columbia.edu}

\altaffiltext{2}{Leiden Observatory, Leiden University, PO Box 9513, 2300 RA, Leiden, The Netherlands}

\altaffiltext{3}{ASTRON, the Netherlands Institute for Radio
  Astronomy, Postbus 2, 7990 AA, Dwingeloo, The Netherlands}

\altaffiltext{4}{Anton Pannekoek Institute for
Astronomy, University of Amsterdam, Science Park 904, 1098 XH
Amsterdam, The Netherlands}

\altaffiltext{5}{Jodrell Bank Centre for Astrophysics, School of Physics and Astronomy, The University of Manchester, Manchester M13 9PL, UK}

\begin{abstract}  
  We present \textit{XMM-Newton} and \textit{Chandra} observations of
  the low-mass X-ray binary XSS J12270--4859, which experienced a
  dramatic decline in optical/X-ray brightness at the end of 2012,
  indicative of the disappearance of its accretion disk. In this new
  state, the system exhibits previously absent orbital-phase-dependent,
  large-amplitude X-ray modulations with a decline in flux at superior
  conjunction. The X-ray emission remains predominantly non-thermal
  but with an order of magnitude lower mean luminosity and
  significantly harder spectrum relative to the previous high flux
  state. This phenomenology is identical to the behavior of the
   radio millisecond pulsar binary PSR J1023+0038 in
  the absence of an accretion disk, where the X-ray emission is
  produced in an intra-binary shock driven by the pulsar wind.  This
  further demonstrates that XSS J12270--4859 no longer has an
  accretion disk and has transformed to a full-fledged eclipsing
  ``redback'' system that hosts an active rotation-powered millisecond
  pulsar. There is no evidence for diffuse X-ray emission associated
  with the binary that may arise due to outflows or a wind nebula. An
  extended source situated 1.5$'$ from XSS J12270--4859 is unlikely to
  be associated, and is probably a previously uncatalogued galaxy
  cluster.
\end{abstract}


\section{INTRODUCTION}
Since the original discovery of millisecond pulsars
\citep[MSPs;][]{Back82} the prevailing theory of their formation has
centered on ``recycling'' by transfer of matter and angular momentum
from a close companion star during a low-mass X-ray binary (LMXB)
phase \citep{Alp82,Rad82}.  The discovery of the first accreting X-ray
MSP, SAX J1808.4--3658 \citep{Wij98}, provided compelling evidence
for this evolutionary sequence \citep[see also][for a review]{Pat12}.
Additional observational support in favor of this hypothesis came with
the discovery of the ``missing link'' radio pulsar PSR J1023+0038
(also known as FIRST J102347.6+003841). Optical observations revealed
an accretion disk in the system in 2001 \citep{Wang09}, which was
absent after 2003 \citep{Thor05} and at the time of the radio pulsar
discovery in 2009 \citep{Arch09}.  The long-suspected evolutionary
connection between LXMBs and radio MSPs was conclusively established when
PSR J1824--2452I in the globular cluster M28 was seen to switch
between rotation-powered (radio) and high-luminosity accretion-powered (X-ray)
pulsations \citep{Pap13}.  

Recent radio, X-ray, and $\gamma$-ray observations revealed that in
2013 June PSR J1023+0038 had undergone another transformation
\citep{Stap14}. Complete cessation of radio pulsations was observed,
accompanied by an extraordinary five-fold increase in the
\textit{Fermi} Large Area Telescope (LAT) flux and enhancement in UV
and X-ray brightness \citep{Pat14,Ten14}.  These recent findings have
revealed that rather than a one-time change from a LMXB to a permanent
radio MSP state, for some systems this phase of neutron star compact
binary evolution involves recurrent switching between the two states.

 In the absence of an accretion disk, PSRs J1023+0038 and PSR
 J1824--2452I belong to the family of so-called ``redbacks''
 \citep[see][and references therein]{Rob11}, namely, eclipsing radio
 MSPs with relatively massive ($\gtrsim$0.2 M$_{\odot}$)
 non-degenerate companions that are (nearly) Roche-lobe filling. These
 objects are distinct from ``black widow'' systems, which are bound to
 very low mass degenerate stars being ablated by the pulsar wind
 \citep[][]{Fru88,Stap96}. Extensive X-ray studies of redbacks in globular
 clusters \citep{Bog05,Bog10,Bog11a} and the field of the Galaxy
 \citep{Arch10,Bog11b,Bog14,Gen13} have revealed that the dominant source of
 X-rays from these binaries is non-thermal radiation with
 $\Gamma\approx 1-1.5$ originating in an intra-binary shock
 \citep{Arons93} generated by the interaction of the pulsar wind and
 matter from the companion star.  The X-ray emission is strongly
 modulated at the orbital period in all redback systems, with a
 decline in flux when the secondary star is between the pulsar and the
 observer. This orbital-phase-dependent X-ray variability appears to
 be a distinguishing characteristic of these peculiar MSP binaries and
 provides a convenient way to identify additional redbacks even in the
 absence of radio pulsations.

The Galactic X-ray source XSS J12270--4859 (1RXS J122758.8--485343)
has posed a mystery since its discovery.  Although initially
classified as a cataclysmic variable (Masetti et al.~2006; Butters et
al.~2008), subsequent multi-wavelength studies (Pretorius~2009; de
Martino et al.~2010; Hill et al.~2011; de Martino et al.~2013; Papitto et
al.~2014) revealed that this system closely resembles a quiescent
LMXB.  XSS J12270--4859 is exceptional in that it was the first LMXB
putatively associated with a $\gamma$-ray source, 1FGL J1227.9--4852
\citep{Hill11}, now known as 2FGL J1227.7--4853 \citep{Nolan12}.
Deep searches have failed to detect radio pulsation from XSS J12270--4859
during its high optical/X-ray flux state \citep{Hill11}.

Prior to 2012 November/December, the X-ray spectrum of XSS
J12270--4859 was well described by a pure powerlaw with $\Gamma=1.7$
and a 0.1--10 keV luminosity of $\sim$$6\times10^{33}$ erg s$^{-1}$
\citep{deM10,deM13}.  The source exhibited occasional intense flares
and peculiar, frequent drops in X-ray flux that are not correlated
with orbital phase. This behavior is reminiscent of the erratic
flux variations observed in PSRs J1023+0038 (Patruno et al.~2014;
Tendulkar et al.~2014; S.~Bogdanov in prep.) and J1824--2452I (Papitto
et al.~2013; Linares et al.~2014) in their accretion disk states,
hinting at a connection with these systems. Recent optical
and X-ray observations revealed that XSS J12270--4859 had undergone a
substantial decline in brightness and no longer exhibits evidence for
an accretion disk (Bassa et al.~2014). The abrupt change to a disk-free
system appears to have occured between 2012 November 14 and 2012
December 21. Given the similarities
with PSR J1023+0038 in both states, XSS J12270--4859 is presently
consistent with hosting a rotation-powered millisecond pulsar.  This
prediction was confirmed with the recent detection of 1.69-millisecond
radio pulsations from the system (Roy et al.~2014).

Here we present recently acquired \textit{XMM-Newton} European
Photon Imaging Camera (EPIC) and \textit{Chandra X-ray Observatory}
Advanced CCD Imaging Spectrometer (ACIS) observations of XSS
J12270--4859 in its new low flux state. These observations confirm
that this source has undergone a metamorphosis to a redback, i.e.~a
compact binary containing a rotation-powered millisecond pulsar.  The
analysis is presented as follows. In \S 2, we provide details on the
observations and data reduction procedures. In \S3, we
present an X-ray variability study. In \S4 we summarize the
phase-averaged and phase-resolved spectroscopy, while in
\S5 we report on an imaging analysis.  We provide a discussion
and conclusions in \S6.

\section{OBSERVATIONS AND DATA REDUCTION}

\subsection{\textit{XMM-Newton} EPIC}
Motivated by the sudden decline in X-ray and optical brightness of
XSS J12270--4859 \citep{Bassa14}, we obtained a target of opportunity
\textit{XMM-Newton} observation on 2013 December 29 (ObsID 0727961401)
for a total exposure time of 38-ks.  The pn \citep{struder01} and two
MOS \citep{turner01} EPIC instruments used the thin optical blocking
filters and were set up in Full Window mode.  The data were processed
using the \emph{XMM-Newton} Science Analysis Software
(SAS\footnote{The \textit{XMM-Newton} SAS is developed and maintained
  by the Science Operations Centre at the European Space Astronomy
  Centre and the Survey Science Centre at the University of
  Leicester.}) version {\tt xmmsas\_20130501\_1901-13.0.0}.  After
applying the recommended flag, pattern, and pulse invariant filters,
the event lists were screened for instances of strong background
flares, which resulted in effective exposures of 31.2, 31.2, and 26.9
ks for the MOS1, MOS2, and pn, respectively.

For the variability and spectral analyses presented in \S3 and \S4,
source counts were extracted from each detector within a circular
region of radius 40$''$, which contains $\sim$90\% of the point source
energy.  The background counts were extracted from three source-free
regions on the same detector chip as the target.  To study the
system's X-ray variability, the photon arrival times were barycentred
using the {\tt barycen} SAS command with the DE405 solar system
ephemeris and assuming the source J2000 coordinates RA=12$^{\rm
  h}$27$^{\rm m}$58$\fs$748,
Dec=$-$48$^{\circ}$53$'$42$\farcs$88. Owing to the coarse read-out
time of the EPIC instruments in Full Window mode (0.73 s and 2.6 s for
the pn and MOS, respectively), it was not possible to search for
pulsations in the millisecond range.

The simultaneous fast photometry U band filter data obtained with the
\textit{XMM-Newton} Optical Monitor (OM) were presented in \citet{Bassa14}.

%
%
\begin{figure}[!t]
\begin{center}
\includegraphics[width=0.45\textwidth]{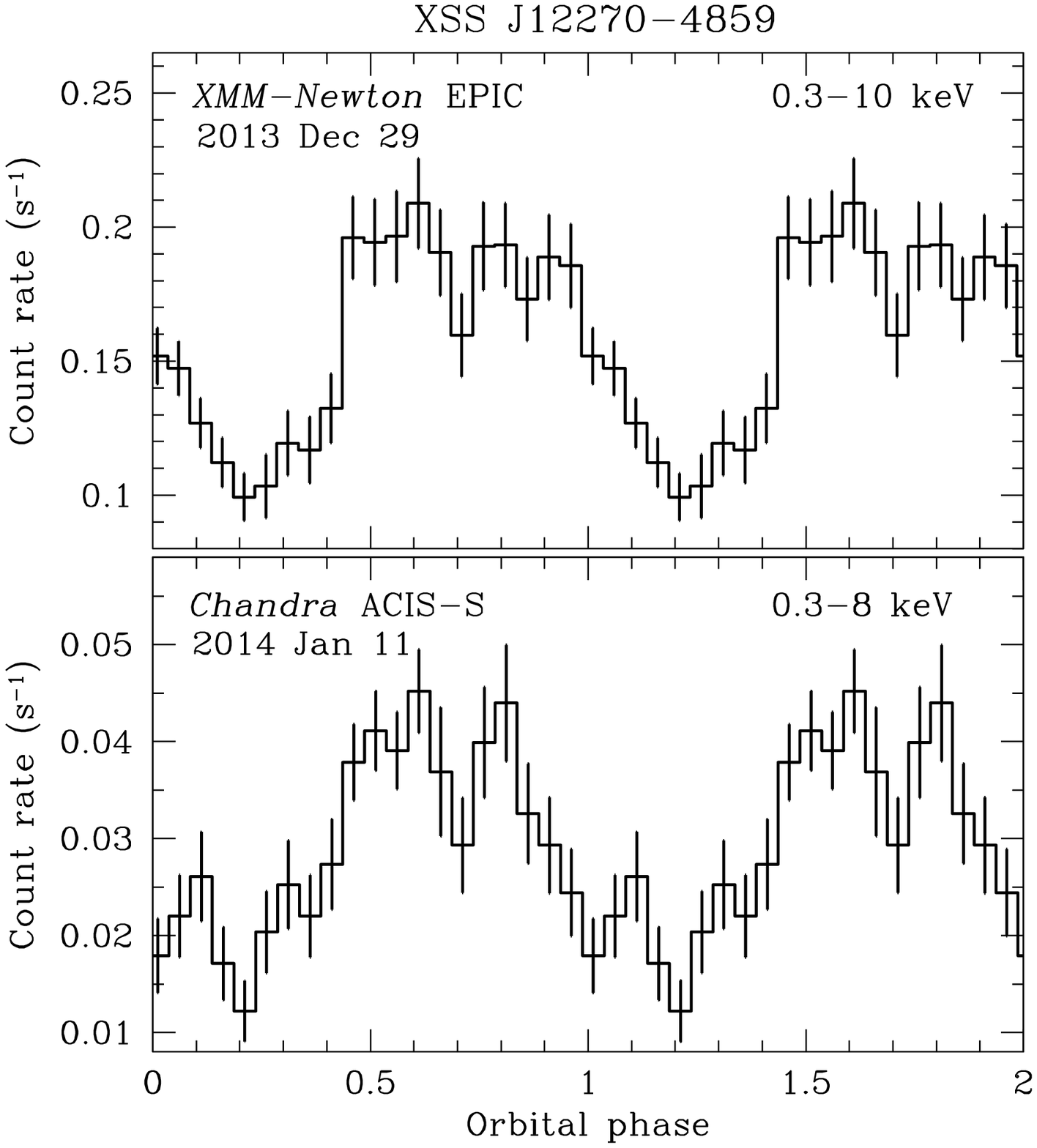}\\
\includegraphics[width=0.45\textwidth]{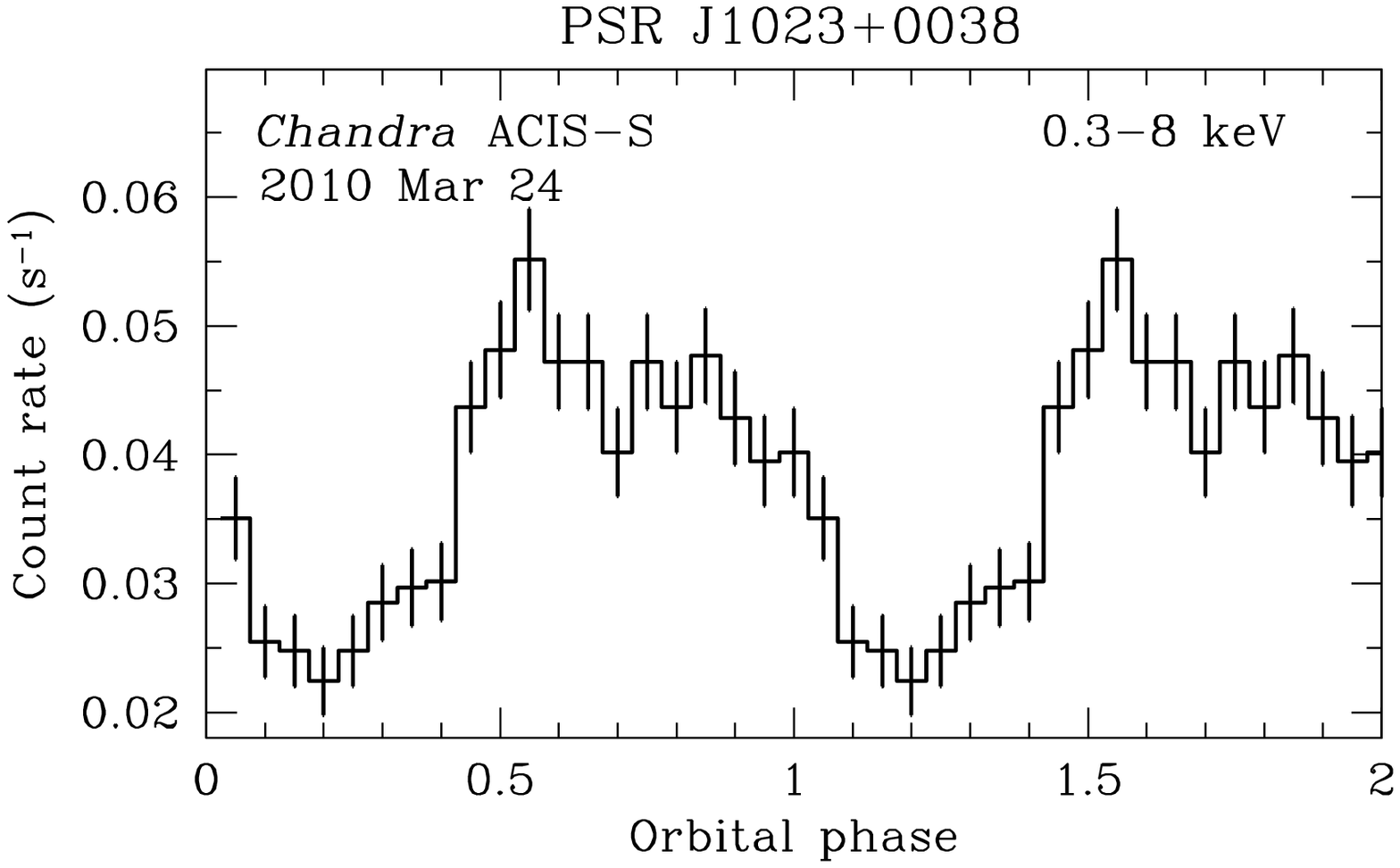}
\end{center}
\caption{Exposure-corrected, background-subtracted lightcurves of XSS
  J12270--4859 from  \textit{XMM-Newton} EPIC
  (\textit{top}) and \textit{Chandra} ACIS-S (\textit{middle})
  observations.  In both cases, the data are folded using the binary
  orbital ephemeris determined in Bassa et al.~(2014). For comparison,
  the bottom panel shows the folded \textit{Chandra} ACIS-S data of
  PSR J1023+0038 in the disk-free radio pulsar state. Two orbital
  cycles are shown for clarity.}
\end{figure}

\subsection{{\it Chandra} ACIS-S}
To look for extended emission around XSS J12270--4859, a target of
opportunity \textit{Chandra} ACIS-S observation was conducted on 2014
January 11 (ObsID 16561) for a duration of 30.2-ks. The best known
position of XSS J12270--4859 was placed at the nominal aim point of
the ACIS S3 chip, which was configured in FAINT telemetry mode.  The
data processing and analysis were performed with CIAO\footnote{Chandra
  Interactive Analysis of Observations, available at
  \url{http://cxc.harvard.edu/ciao/}.} 4.6 \citep{Fruscione06} and
CALDB 4.5.9 using standard procedures.

For the variability and spectroscopic studies, we extracted photons
within 2$\arcsec$ of the source.  For the variability study the event
times were translated to the solar system barycenter assuming the
DE405 JPL solar system ephemeris and the same position for XSS
J12270--4859 used for the \textit{XMM-Newton} data.  The 3.2-s
read-out for ACIS-S prevented a search for pulsations.  To produce
spectra suitable for analysis, the extracted source counts in the
0.3--8 keV range were grouped so as to ensure at least 15 counts per
energy bin. The background was taken from three source-free circular
regions near the target.

\section{X-ray Orbital Variability}
Based on the refined binary orbital period ($P_b=6.913\pm0.002$ hours)
and reference epoch of the ascending node (HJD $2456651.026\pm0.002$
corresponding to orbital phase $\phi_b=0$) of XSS J12270--4859
obtained in \citet{Bassa14} we have folded the barycentered
\textit{XMM-Newton} and \textit{Chandra} point source X-rays at the
binary period. Both observations span slightly more than one binary
orbit. Large-amplitude variability correlated with binary phase is
immediately evident from Figure 1. This behavior is in stark contrast
with the \textit{XMM-Newton} observations from 2009 and 2011
\citep{deM10,deM13} where no orbital dependence on the X-ray flux is
seen and only random, aperiodic variability is present. To formally
establish the statistical significance of the variability, we use a
Kuiper test \citep{Pal04}, which accounts for the non-uniform exposure
across the orbit of the folded lightcurves and is not dependent on the
choice of binning of the lightcurve, making it a better-suited choice
than the Kolmogorov-Smirnov and $\chi^2$ tests for this
purpose. Applying this test to the folded, unbinned lightcurves,
yields $7\times10^{-41}$ (13.3$\sigma$) and $7\times10^{-15}$
(7.7$\sigma$) probabilities for the \textit{XMM-Newton} and
\textit{Chandra} data, respectively, that photons drawn from a
constant distribution would exhibit this level of non-uniformity.

For comparison, the folded lightcurve of PSR J1023+0038 from a 83-ks
observation obtained on 2010 March 24 with \textit{Chandra} ACIS-S
(ObsID 11075), when the binary was in an accretion disk-free state
\citep[see][for further details]{Bog11b}, is shown in the bottom panel
of Figure 1. The close similarities between the lightcurves both in
morphology and phase alignment are quite striking, further indication that
the two systems are close analogs.

\begin{deluxetable}{lccc}
\tabletypesize{\small} 
\tablecolumns{4} 
\tablewidth{0pc}
\tablecaption{Results of Joint \textit{XMM-Newton/Chandra} Spectral Fits for XSS J12270--4859.}

\tablehead{\colhead{} & \colhead{Total} &
  \colhead{$\phi_{b,1}$} & \colhead{$\phi_{b,2}$} \\
  \colhead{Model\tablenotemark{a}} & \colhead{ } &
  \colhead{$(0.0-0.5)$} & \colhead{$(0.5-1.0)$}}
 \startdata
\textbf{PL} & 	&	&		\\
\hline
$N_{\rm H}$ ($10^{20}$ cm$^{-2}$) 	  & $6.1\pm1.1$	& $3.6^{+1.6}_{-1.5}$	& $5.2^{+0.9}_{-0.5}$	\\
$\Gamma$	          & $1.20\pm0.04$	& $1.19\pm0.06$	& $1.11\pm0.04$ \\
$F_X$ ($0.1-10$ keV)\tablenotemark{c} & $4.6\pm0.1$  	& $3.7\pm0.1$	& $6.0\pm0.2$	\\
$\chi^2_{\nu}$/dof       & $0.94/216$ & $0.90/100$	& $0.93/118$	\\
\hline
\textbf{PL + NSA\tablenotemark{b}}	& 	&	&	\\
\hline
$N_{\rm H}$ ($10^{20}$ cm$^{-2}$)	  & $14.5^{+7.6}_{-7.4}$  & $1.8^{+7.2}_{-1.7}$ & $10.2^{+7.2}_{-6.1}$	\\
$\Gamma$	        & $1.16^{+0.07}_{-0.08}$	& $0.98\pm0.2$	& $1.04^{+0.10}_{-0.07}$ 	\\
$T_{\rm eff}$ ($10^6$ K)       & $0.82^{+0.83}_{-0.23}$	& $2.73^{+1.40}_{-1.97}$ & $1.13^{+1.46}_{-0.51}$ 	\\
$R_{\rm eff}$ (km)      &  $0.22^{+5.17}_{-0.22}$        & $<0.19$ 	& $<1.7$ 	\\
Thermal fraction\tablenotemark{d}      &  $0.04\pm0.05$ & $0.06\pm0.03$ 	& $0.07\pm0.03$ 	\\
$F_X$ ($0.1-10$ keV)\tablenotemark{c}	& $4.8\pm0.1$ 	& $3.7\pm0.1$ & $6.5\pm0.2$ 	\\
$\chi^2_{\nu}/$dof       & $0.91/214$ & $0.90/98$ & $0.92/116$  
\enddata 

\tablenotetext{a}{All uncertainties quoted are 1$\sigma$.}
\tablenotetext{b}{For the {\tt nsa} model, a $M=1.4$ M$_{\odot}$, $R=10$ km, and $B=0$
  neutron star is assumed. The effective emission radius as measured
  at the stellar surface, $R_{\rm eff}$, is calculated assuming the distance range of $1.4-3.6$ kpc.}
\tablenotetext{c}{Unabsorbed X-ray flux ($0.1-10$ keV) in units of
  $10^{-13}$ erg cm$^{-2}$ s$^{-1}$.}
\tablenotetext{d}{Fraction of unabsorbed flux from the thermal component in the 0.1--10 keV band.} 

\end{deluxetable}

%
%
\begin{figure}[!t]
\begin{center}
\includegraphics[width=0.45\textwidth]{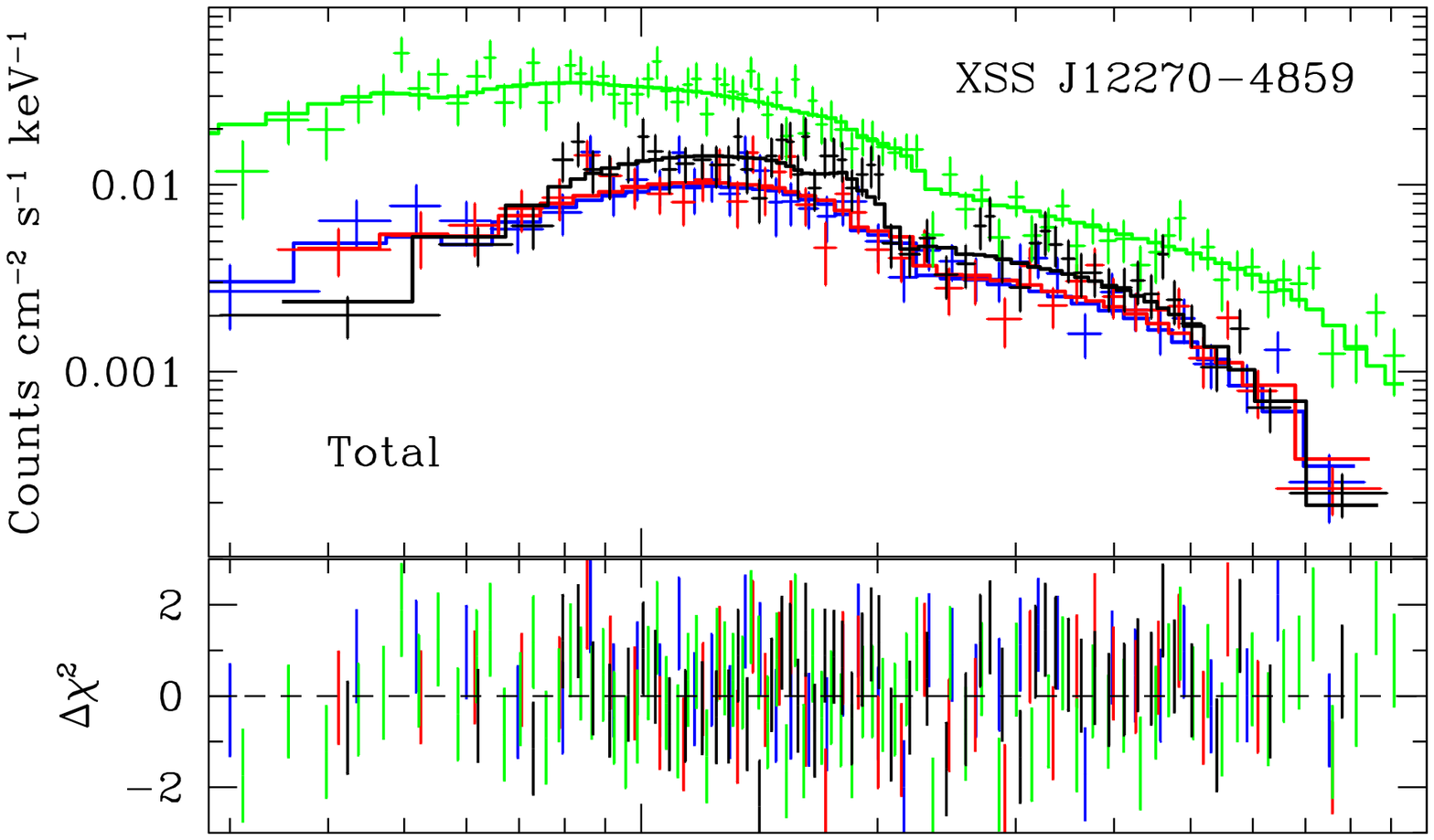}\\
\includegraphics[width=0.45\textwidth]{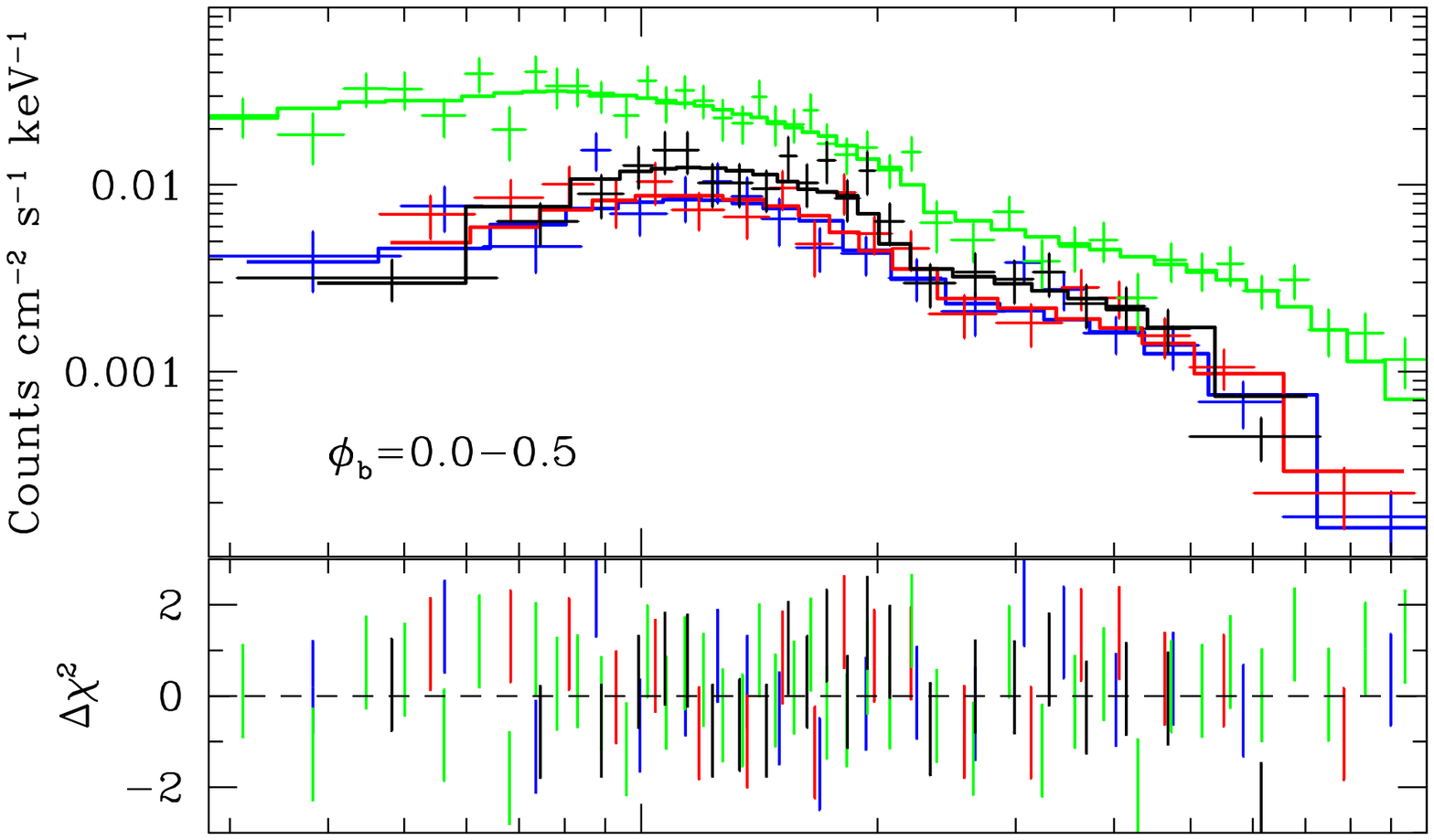}\\
\includegraphics[width=0.45\textwidth]{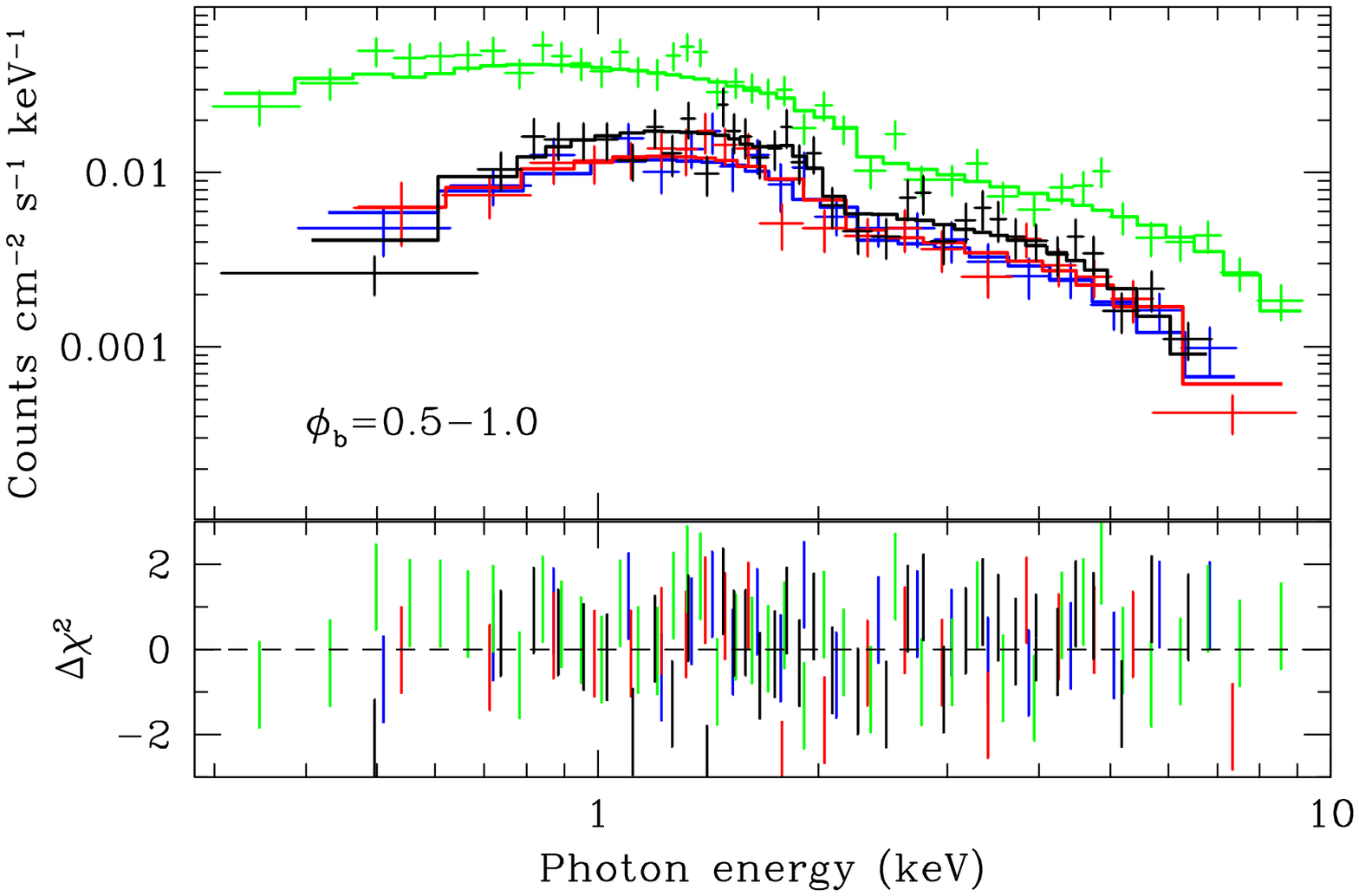}
\end{center}
\caption{\textit{XMM-Newton} EPIC MOS1/2 (\textit{red}/\textit{blue}),
  EPIC pn (\textit{green}), and \textit{Chandra} ACIS-S
  (\textit{black}) phase-averaged X-ray spectra (\textit{top}) and for
  orbital phases $\phi_b=0.0-0.5$ (\textit{middle}) and
  $\phi_b=0.5-1.0$ (\textit{bottom}), fitted with an absorbed
  power-law. The smaller panels show the best-fit residuals in terms
  of $\sigma$ with error bars of size one. For the best fit
  parameters, see text and Table 1.}
\end{figure}

\section{X-ray Spectroscopy}
\subsection{Phase-averaged Spectroscopy}

The spectroscopic analysis was carried out in XSPEC\footnote{Available
  at \url{http://heasarc.nasa.gov/docs/xanadu/xspec/index.html}.}
12.7.1 \citep{Arnaud96}. Preliminary fits of the individual
\textit{XMM-Newton} and \textit{Chandra} spectra produced consistent
results so all fits were performed jointly on both data sets to obtain
the best constraints on the X-ray spectral properties of XSS
J12270--4859.  Fitting the phase-averaged spectra with an absorbed
power-law results in a statistically acceptable fit (see Table 1 and
top panel of Figure 2), with $\chi^2_{\nu}=0.94$ for 216 d.o.f., while
a single or even a two-temperature neutron star atmosphere thermal
spectrum does not yield a satisfactory fit, with $\chi^2_{\nu}=2.6$
for 216 d.o.f. and $\chi^2_{\nu}=1.6$ for 214 d.o.f.,
respectively. Since XSS J12270--4859 currently hosts an active
rotation-powered pulsar, X-ray-emitting magnetic polar caps heated by
a back flow of particles from the pulsar magnetosphere may be
present, as seen in many radio MSPs \citep[see,
  e.g.,][]{Zavlin06,Bog06}. Therefore, in addition to the dominant
non-thermal emission there may be a non-negligible contribution from
the pulsar polar caps.  Including the non-magnetic neutron star
hydrogen atmosphere {\tt nsa} model \citep{Zavlin96} component results
in an acceptable fit but the parameters of the thermal component are
poorly constrained. The thermal flux contribution is consistent with
zero even at a 1$\sigma$ confidence level and the upper limit on the
thermal fraction is $\le$9\%.

For both the power-law and composite models, the derived hydrogen
column densities along the line of sight, $N_{\rm H}$, are in
agreement with the value measured through the Galaxy of
$\sim$$1\times10^{21}$ cm$^{-2}$ in the direction of XSS
J12270--4859 \citep{Kalb05}.  Assuming the distance range of $1.4-3.6$
kpc \citep{deM13}, the derived time-averaged unabsorbed fluxes imply
an X-ray luminosity of $(1-7)\times10^{32}$ erg s$^{-1}$ in the
0.1--10 keV interval, an order of magnitude fainter than the average
luminosity in the accretion disk dominated state.

%
\begin{figure}[t]
\begin{center}
\includegraphics[width=0.47\textwidth]{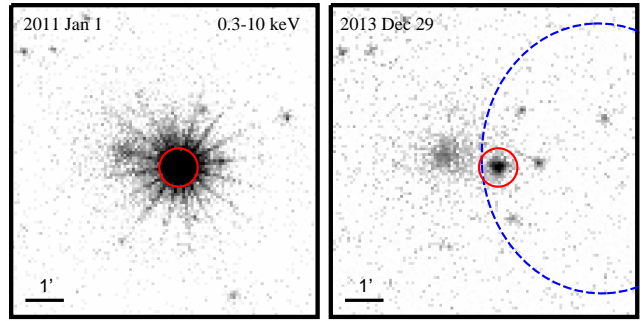}
\end{center}
\caption{\textit{XMM-Newton} EPIC MOS1+2 mosaic images in the 0.3--10
  keV band from 2011 Jan 1 (\textit{left}), during the high state, and
  2013 Dec 29 (\textit{right}), in the current low state of XSS
  J12270--4859. The red circle of radius 40$''$ is centered on the
  position of XSS J12270--4859. The 95\% confidence error ellipse of
  2FGL J1227.7--4853 is shown with the dashed blue line. The grayscale
  shows intensity increasing logarithmically from white to
  black. North is up and east is to the left.}
\end{figure}

\subsection{Orbital-Phase-resolved Spectroscopy}
To look for changes in the X-ray spectrum as a function of the binary
orbital phase, we divided the data in two equal phase portions:
$\phi_b=0.0-0.5$ (encompassing the X-ray minimum) and $\phi_b=0.5-1.0$
based on the orbital ephemeris from \citet{Bassa14}. The spectral fits
are summarized in columns 3 and 4 of Table 1. An absorbed power-law
model yields good fits for both phase intervals (Figure 2) with
similar values for the spectral photon index $\Gamma$, implying no
appreciable spectral variability over the orbit despite the factor of
1.5--1.7 difference in flux. As with the phase-averaged analysis, the
addition of an {\tt nsa} model component also results in acceptable
fits. For both orbital phase intervals any thermal emission
contributes $\lesssim$10\% to the $0.1-10$ keV flux.  The {\tt nsa}
temperature and emission radii are poorly constrained but are in
general agreement with those observed for thermally-emitting MSPs
\citep[e.g.,][]{Bog06,Zavlin06,Bog09}.

\section{X-ray Imaging Analysis}
Even in its current low state, XSS J12270--4859 still remains the brightest
X-ray source in its field. The substantially lower flux allows a close
examination of the spatial distibution of emission around the binary
for the first time. For XSS J12270--4859, this is especially
interesting since a currently active rotation-powered pulsar wind may,
in principle interact with circum-binary material that was expelled
during the LMXB state via outflows. Moreover, the resulting wind
nebula may only be visible for a brief period after the state
transition since presumably the material would eventually be swept up
by the pulsar wind.

In the most recent images, a diffuse source situated $\sim$1.5$'$
east-north-east of XSS J12270--4859 is clearly visible (Figures 3 and
4), which can only barely be seen in the \textit{XMM-Newton} EPIC
MOS1+2 image from the high state (2011 January 1) due to the much
brighter wings of the point spread function of XSS J12270--4859. The
centroid of the diffuse source, which is $\sim$$40-60$$\arcsec$ in
extent, is at approximately RA=12$^{\rm h}$28$^{\rm m}$07$\farcs$5 and
Dec=$-$48$^{\circ}$53$'$24$''$ (J2000).  To establish if the two
objects are associated, we looked for emission connecting XSS
J12270--4859 with the extended region by comparing the count rates
around XSS J12270--4859 in the direction towards and away from the
diffuse source. This was accomplished by considering a set of adjacent
8$''$ by 16$''$ rectangular regions aligned along the axis connecting
the centroid of the diffuse source and XSS J12270--4859 (see Figure
4). The size was chosen so as to ensure that a sufficient number of
photons are collected in each region. Comparing the distribution of
counts towards and away from the diffuse region (Figure 5) shows that
the emission from the diffuse source does not extend to the position
of XSS J12270--4859, suggesting no physical connection between the two
X-ray sources.

In addition, if the two objects are in fact associated, the angular
separation of $\sim$1.5$'$ would translate to a physical separation of
$0.6-1.6$ pc, and the size of the diffuse region to $0.3-1$ pc, for an
assumed distance range of $1.4-3.6$ kpc.  These physical scales are
much larger than what is observed in other accreting systems with
outflows, such as the microquasars XTE J1550--564 \citep{Kaa03,Tom03}
and H1743--322 \citep{Cor05}. Therefore, it is unlikely that the
extended emission is a product of an outflow from the
binary. Likewise, a pulsar wind nebula (PWN) can be ruled out because
XSS J12270--4859 is not embedded in the nebular emission, which also
does not resemble a bow shock or a trailing tail \citep[see,
  e.g.,][for an overview of X-ray PWNe]{Kar08}.  These factors
indicate that an association between the binary and the diffuse source
is highly unlikely.  It is more likely that the extended object is
simply a chance superposition and is unrelated to the
binary. Examination of VLT/FORS2 images ($3\times15$\,s exposure in
$R$-band obtained on 2010 February 10) show a bright source with
morphology resembling that of a galaxy located at RA$=$12$^{\rm
  h}$28$^{\rm m}$07$\fs$427, Dec.$=$$-$48$^{\circ}$53$'$23$\farcs$561
(see bottom panel of Figure 4), which is coincident with the centroid
of the diffuse X-ray region and is surrounded by several fainter
objects.  Therefore, this is likely the central galaxy of a previously
uncatalogued galaxy cluster, which we designate as J122807.4--48532,
with the diffuse X-ray emission originating from the intra-cluster
medium \citep{Cav76}. Additional optical observations are required to
conclusively establish this scenario.

%
\begin{figure}[t]
\begin{center}
\includegraphics[width=0.46\textwidth]{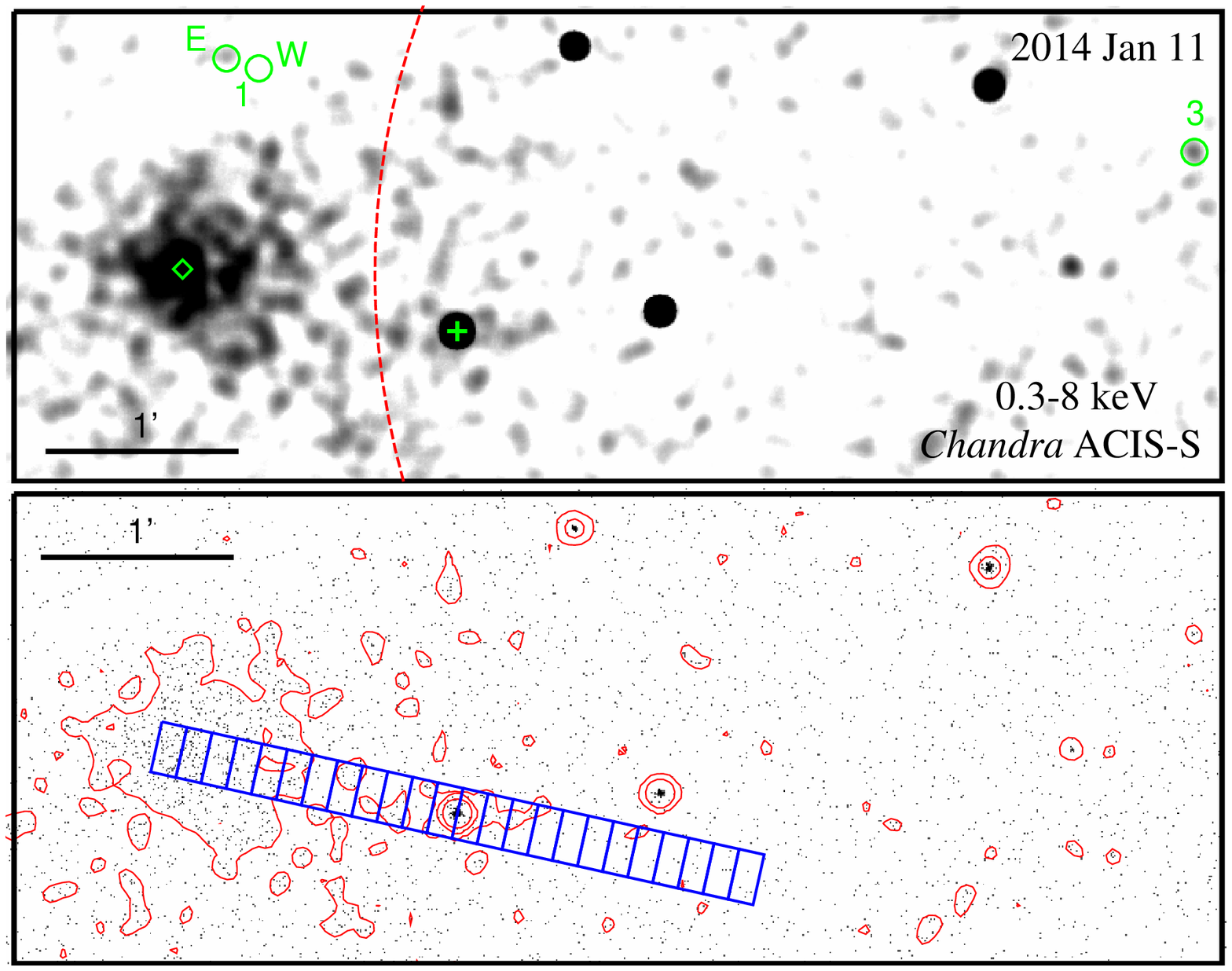}
\includegraphics[width=0.455\textwidth]{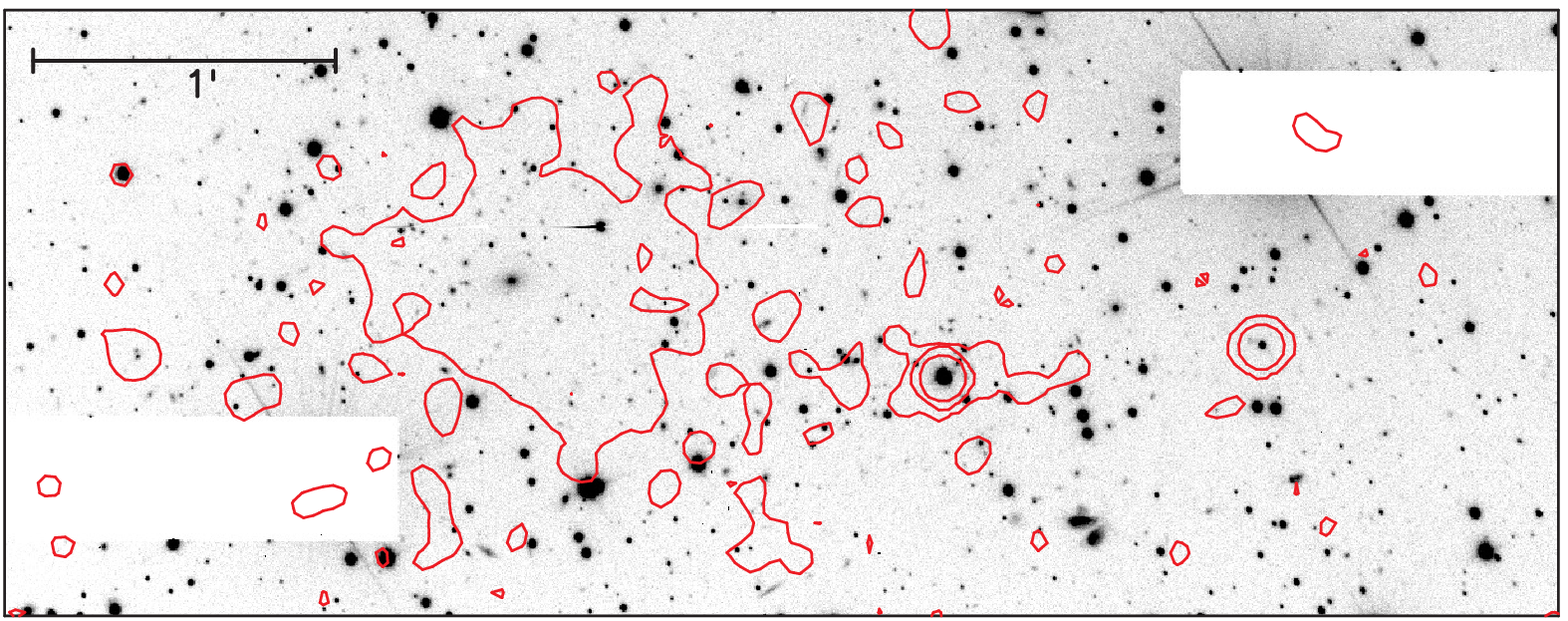}
\end{center}
\caption{\textit{Top}: \textit{Chandra} ACIS-S image in the 0.3--8 keV
  band from 2014 Jan 11 smoothed with a two-dimensional Gaussian
  kernel of width 5$''$. The green $+$ marks the position of XSS
  J12270--4859, the green diamond is centered on the optically
  identified galaxy (see text), and the green circles mark the ATCA
  radio sources 1 and 3 reported by Hill et al.~(2011). The 95\%
  confidence error ellipse of 2FGL J1227.7--4853 is shown with the
  dashed red line. The grayscale shows intensity increasing
  logarithmically from white to black. \textit{Middle}: The same image
  binned at the intrinsic detector resolution of 0.5$''$. The red
  contour levels were obtained from the smoothed image in the top
  panel.  The blue strip of rectangles shows the regions used to
  extract the count rates shown in Figure 5. Pixel randomization has
  been removed from the pipeline processing. \textit{Bottom}:
  VLT/FORS2 image of the field around XSS J12270--4859 with the same
  X-ray contours shown in the middle panel. Note the extended optical
  source near the middle of the diffuse X-ray region surrounded by a
  number of fainter sources.}
\end{figure}

The 0.5$''$ angular resolution of the ACIS-S image enables an
investigation of any nebular emission in the immediate vicinity of the
binary (within $\sim$20$''$).  For this purpose, we have simulated 10
observations of XSS J12270--4859 using ChaRT\footnote{The Chandra Ray
  Tracer, available at
  \url{http://cxc.harvard.edu/soft/ChaRT/cgi-bin/www-saosac.cgi}} and
MARX 4.5\footnote{Available at
  \url{http://space.mit.edu/cxc/marx/index.html}.} assuming the
parameters of the ACIS-S exposure and the best-fit non-thermal model
reported in \S 4. The mean of the simulated observations was
subtracted from the real data to look for excess emission around the
binary. However, as apparent from Figure 6, the radial profile of the
X-ray emission from XSS J12270--4859 is fully consistent with that of
a point source.

We note that this diffuse X-ray source has no radio counterpart in the
ATCA image presented by \citet{Hill11}. On the other hand, the eastern
component of the radio source designated by \citet{Hill11} as 1, as
well as their source 3, are coincindent with very faint X-ray sources
(marked in Figure 4). Since these sources are at the threshold of
detection in the \textit{Chandra} image, their spectral properties
cannot be well constrained.

%
%
\begin{figure}[t]
\begin{center}
\includegraphics[width=0.44\textwidth]{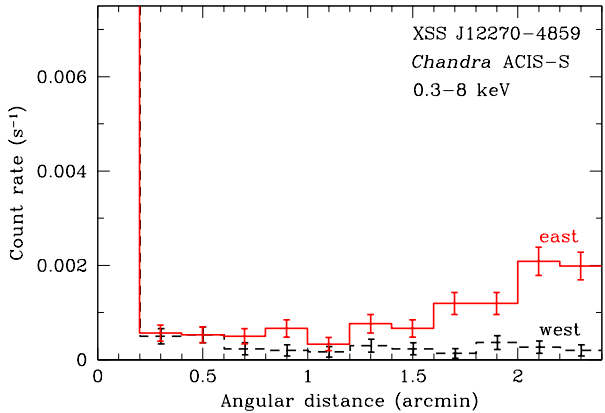}
\end{center}
\caption{Distribution of count rates in the directions towards (east)
  and away (west) from the nearby diffuse source relative to the
  centroid of XSS J12270--4859. The count rates were extracted from
  the rectangular regions depicted in Figure 4.}
\end{figure}

%
\begin{figure}[t]
\begin{center}
\includegraphics[width=0.44\textwidth]{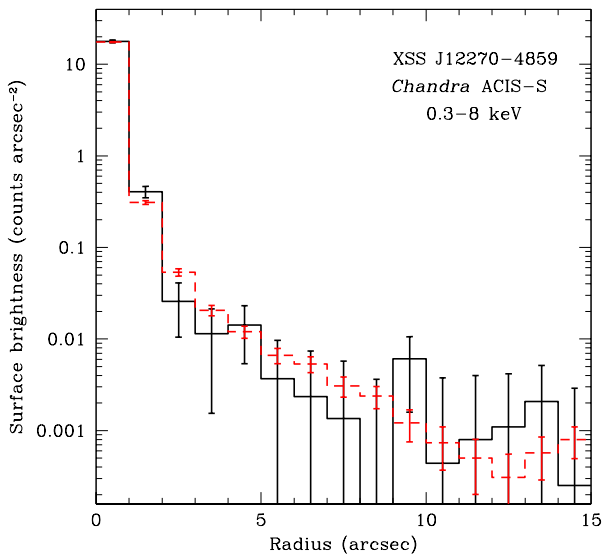}
\end{center}
\caption{Background-subtracted radial profile of XSS J12270--4859 from
  the \textit{Chandra} ACIS-S observation (\textit{black})
  and a synthetic point spread function generated using ChaRT and MARX
  (\textit{red dashed line}).}
\end{figure}

\section{DISCUSSION AND CONCLUSIONS} 

As reported in Bassa et al.~(2014), the recent decline in optical and X-ray
brightness and the disappearance of the previously prominent optical
emission lines are indicative of the disappearance of the accretion disk
in XSS J12270--4859. The \textit{XMM-Newton} and \textit{Chandra}
observations of XSS J12270--4859 in its new low flux state presented
herein reveal further information regarding this binary:

1.~Previously absent large-amplitude modulation at the binary orbital
period is now present.

2.~The time-averaged X-ray luminosity of $\sim$$(1-7)\times 10^{32}$
erg s$^{-1}$ is smaller by an order of magnitude relative to the
accretion disk dominated state.

3.~The predominantly non-thermal spectrum with spectral photon index
of $\Gamma=1.2$ is significantly harder than the $\Gamma=1.7$ observed
in the high flux state.

4.~The spectral index of the power-law emission appears not to vary
substantially even though the flux changes by a factor of
$\sim$1.5--1.7 over an orbital period.
 
These X-ray properties, as well as the optical properties described in
\citet{Bassa14}, are identical to what is observed from ``redback''
radio MSPs, including PSRs J1023+0038 \citep{Arch09,Arch10,Bog11b},
J1723--2837 \citep{Faulk04,Craw13,Bog14}, J2215+5135
\citep{Hes11,Gen13} in the field of the Galaxy, as well as PSR
J0024--7204W in the globular cluster 47 Tuc
\citep{Camilo00,Freire03,Bog05}, PSR J1740--5340 in NGC 6397
\citep{DAm01,Bog10}, and PSR J1824--2452H in M28 \citep{Bog11a,Pal10}.
Therefore, the results obtained herein, and the recent radio pulsar
discovery by \citet{Roy14} further demonstrate that XSS J12270--4859
is a redback, i.e. a compact binary with an active rotation-powered
pulsar and no accretion disk.

The non-thermal X-rays from redback systems are generated in an
intra-binary shock formed by the interaction of the pulsar wind with
material from the close companion star \citep{Arons93}. As shown in
\citet{Bog11b}, the decline in X-ray flux at $\phi_b\approx0.25$ seen
in PSR J1023+0038 can be reproduced by a simple geometric model in
which the intra-binary shock is partially occulted by the bloated
secondary star. In this scenario, the depth and phase extent of the
X-ray eclipse require that the shock be situated at or very near the
surface of the secondary star on the side exposed to the pulsar wind.

There is no indication of extended X-ray emission associated with XSS
J12270--4859 that could arise due to outflows in the LMXB phase or the
interaction of the pulsar wind with the ambient medium.  An adjacent
diffuse source is likely not associated with the binary and is
probably a background galaxy cluster. Given that only two MSPs, PSRs
B1957+21 \citep{Stap03} and J2124--3358 \citep{Hui06} show evidence
for X-ray bow shocks, the absence of nebular emission associated with
XSS J12270--4859 does not imply the lack of a pulsar wind.

The lack of knowledge of the orbital separation, the companion mass,
and the pulsar spin-down luminosity ($\dot{E}$) for XSS J12270--4859,
do not allow meaningful constraints on the physics of the intrabinary
shock. Nevertheless, given the nearly identical X-ray properties
compared to other redbacks it is possible to place crude constraints
on the energetics of the MSP in XSS J12270--4859. In particular, if we
consider that the shock luminosity at X-ray maximum is
$\sim$$2\times10^{-3}\dot{E}$ for PSRs J1023+0038
\citep{Arch10,Bog11b} and J1723--2837 \citep{Bog14}, the implied
spin-down of the pulsar in XSS J12270--4859 is $(0.7-5)\times10^{35}$
erg s$^{-1}$ assuming a distance range $1.4-3.6$ kpc. This suggests
that the pulsar in XSS J12270--4859 may be among the small number of
energetic MSPs with $\dot{E}\gtrsim10^{35}$ erg s$^{-1}$.  The
0.1--100 GeV $\gamma$-ray luminosity of 2FGL J1227.7--4853, the
putative $\gamma$-ray counterpart of XSS J12270--4859, of
$L_{\gamma}=(0.8-5)\times10^{34}$ erg s$^{-1}$ provides an independent
hard lower limit on the $\dot{E}$ of the pulsar, if we assume a
$\gamma$-ray conversion efficiency of 100\%. For efficiencies typical
of the MSP population \citep[$\lesssim$50\%; see Table 10
  in][]{Abdo13}), this yields $\dot{E}\gtrsim10^{35}$ erg s$^{-1}$, in
general agreement with the limit obtained from the X-ray luminosity.
Continued timing of the radio pulsar would allow a determination of
the true $\dot{E}$ of this pulsar.

The redback nature of the XSS J12270--4859 binary determined from the
X-ray analysis presented above implies that the pulsar should undergo
radio eclipses of frequency-dependent duration around superior
conjunction. These eclipses are caused by intra-binary plasma,
presumably emanating from the companion star.  For most redback and
black widow systems that are detectable in radio pulsations, the radio
emission is eclipsed for $\lesssim$50\% of the orbit, although there
is indication that for some systems this fraction can be close to
100\% \citep[see][for the case of PSR J1311--3430]{Ray13}.  The
non-detections with the Parkes telescope \citep{Bassa14} and only a
single, brief detection with the GMRT \citep{Roy14} despite deep
pulsation searches, are indicative of a high radio eclipse fraction
for XSS J12270--4859 that likely varies from orbit to orbit.  The
large $\dot{E}$ estimated above is consistent with this scenario,
since an energetic wind would drive off material from the surface of
companion at a higher rate than an MSP with a lower $\dot{E}$. This
may result in occasional severe enshrouding by this stripped
material, causing the pulsar to be eclipsed at radio frequencies for a
larger fraction of the orbit. This would make XSS J12270--4859 a
so-called ``hidden'' MSP, as postulated by \citet{Tav91}.

The X-ray properties of XSS J12270--4859 in both the present disk-free
radio pulsar state and the accretion disk-dominated state
\citep[see][]{deM10,deM13} are virtually identical to those of PSRs
J1023+0038 \citep{Arch10,Bog11b,Pat14} and J1824--2452I
\citep{Pap13,Lin14}. This remarkable consistency in X-ray
characteristics provides a powerful discriminant for identifying more
such MSP binaries, especially ones that are heavily enshrouded and
hence cannot be identified via radio pulsation searches.

\acknowledgements A.M.A.~and J.W.T.H.~acknowledge support from a Vrije
Competitie grant from NWO. C.B. acknowledges support from ERC Advanced
Grant ``LEAP'' (227947, PI: Michael Kramer). J.W.T.H. and
A.P. acknowledge support from NWO Vidi grants. J.W.T.H.~also
acknowledges funding from an ERC Starting Grant ``DRAGNET'' (337062).
A portion of the results presented was based on observations obtained
with \textit{XMM-Newton}, an ESA science mission with instruments and
contributions directly funded by ESA Member States and NASA. The
scientific results reported in this article are based in part on
observations made by the \textit{Chandra X-ray Observatory}. This
research has made use of the NASA Astrophysics Data System (ADS),
software provided by the Chandra X-ray Center (CXC) in the application
package CIAO.

Facilities: \textit{XMM,CXO}

\end{document}